# $L^2$ series solutions of the Dirac equation for power-law potentials at rest mass energy

A. D. Alhaidari

*Physics Department, King Fahd University of Petroleum & Minerals, Dhahran 31261, Saudi Arabia*
email: haidari@mailaps.org

We obtain solutions of the three dimensional Dirac equation for radial power-law potentials at rest mass energy as an infinite series of square integrable functions. These are written in terms of the confluent hypergeometric function and chosen such that the matrix representation of the Dirac operator is tridiagonal. The "wave equation" results in a three-term recursion relation for the expansion coefficients of the spinor wavefunction which is solved in terms of orthogonal polynomials. These are modified versions of the Meixner-Pollaczek polynomials and of the continuous dual Hahn polynomials. The choice depends on the values of the angular momentum and the power of the potential.

PACS numbers: 03.65.Pm, 03.65.Ge, 02.30.Gp

Keywords: Dirac equation, power-law potential, zero energy, 3-term recursion relations, tridiagonal representations, Pollaczek polynomials, continuous dual Hahn polynomials.

## I. INTRODUCTION

Exact solutions of the wave equation at zero energy attracted attention [1-8] motivated in part by developments in supersymmetric quantum mechanics [9] and in the search for conditionally exactly [10,11] and quasi exactly [12,13] solvable problems. From mathematical point of view these solutions are interesting since they form, by definition, quasi exactly solvable systems due to the fact that they are solvable only for $E = 0$. Moreover, and despite common intuition brought about by wide familiarity with the Coulomb problem, some of these solutions are square integrable and correspond to bound or unbound states [1-3,6,14]. Furthermore, these solutions are very valuable for zero energy limit calculations in various fields of physics. Such examples are in the study of loosely bound systems, as well as, in scattering length and coupling parameters calculation.

It is elementary to note that there exists any number of potentials for the Schrödinger equation with just one known zero energy eigenstate. This can be seen by noting that the Schrödinger equation $-\chi'' + V\chi = 0$ gives the potential $V = U^2 - U'$, where $U = -\chi'/\chi$ and $\chi$ is nodeless. In unbroken supersymmetry [9], $U^2 \pm U'$ are known as the two superpartner potentials sharing the same energy eigenvalues (i.e., they are isospectral) except for the zero energy ground state, which belongs only to $U^2 - U'$. For our present treatment, however, symmetry is imposed on the solution space of the problem from the outset resulting in a special class of solutions.

In a previous paper [6], we made an attempt to solve the relativistic problem at $E = 0$, where $E$ is the nonrelativistic energy and the potential function was taken as a power-law type potential. However, the success of our findings was very limited. It turns out, as we shall see, that the reason for this short coming is due to the stringent constraint that was placed on the solution space of the problem. This can be explained as follows. Let the spinor wave function $\chi$ be an element in a linear vector space with a complete basis set $\{\psi_n\}_{n=0}^{\infty}$. Then, we can expand it as $|\chi(\vec{r},\varepsilon)\rangle = \sum_n f_n(\varepsilon)|\psi_n(\vec{r})\rangle$, where $\vec{r}$ is the configuration space coordinate and $\varepsilon$ is the relativistic energy. In [6], our search for bases was limited to those that carry diagonal matrix representations for the Hamiltonian $H$ at rest mass energy. That is, we required $H|\psi_n\rangle = \varepsilon_n|\psi_n\rangle = \pm|\psi_n\rangle$. Consequently, we could obtain a solution only for the case when $n = 0$. In this work, however, we relax this constraint by searching for square integrable bases that could support a tridiagonal matrix representation of the wave operator. That is, the action of the wave operator on the elements of the basis is allowed to take the general form $(H-\varepsilon)|\psi_n\rangle \sim |\psi_n\rangle + |\psi_{n-1}\rangle + |\psi_{n+1}\rangle$ such that

$$\langle \psi_n | H - \varepsilon | \psi_m \rangle = (A_n - y)\delta_{n,m} + B_n \delta_{n,m-1} + B_{n-1}\delta_{n,m+1} \tag{1.1}$$

where $y$ and the coefficients $\{A_n, B_n\}_{n=0}^{\infty}$ are real and, in general, functions of the energy, angular momentum, and potential parameters. Therefore, the matrix representation of the wave equation, which is obtained by expanding $|\chi\rangle$ as $\sum_m f_m|\psi_m\rangle$ in $(H-\varepsilon)|\chi\rangle = 0$ and projecting on the left by $\langle \psi_n|$, results in the following three-term recursion relation

$$y f_n = A_n f_n + B_{n-1} f_{n-1} + B_n f_{n+1} \tag{1.2}$$

Consequently, the problem translates into finding solutions of the recursion relation for the expansion coefficients of the wavefunction. This will be solved easily and directly by correspondence with those for well known orthogonal polynomials. It is obvious that the solution of (1.2) is obtained modulo an overall factor which is a function of the physical parameters of the problem but, otherwise, independent of $n$. The uniqueness of the solution is achieved by the requirement of normalizability of the wavefunction, $\langle \chi | \chi \rangle = 1$. Moreover, the matrix wave equation (1.1) shows that the diagonal representation, which we have obtained in [6], is a special case that could easily be obtained by the requirement:

$$B_n = 0, \text{ and } A_n - y = 0. \tag{1.3}$$

Thus, the solution space with these constraints could be extremely limited.

The paper is organized as follows: in the following section, we formulate the problem by writing the three dimensional Dirac equation with non-minimal coupling to a four-potential. Spherical symmetry is imposed and we consider the case where the "even component" of the relativistic potential vanishes while the "odd component" is a power-law radial potential. We exclude the well-known cases where the potential corresponds, for example, to the Dirac-Oscillator problem or the free case. The main results are obtained in Sec. III where we select an $L^2$ spinor basis that supports a tridiagonal matrix representation for the Dirac wave operator. This results in a three-term recursion relation for the expansion coefficients of the wavefunction. The solution of this recursion is given in terms of either a "Hyperbolic" Meixner-Pollaczek polynomial or a "modified" continuous dual Hahn polynomial depending on the values of the physical parameters. The paper concludes with a short discussion about the negative energy solutions and the diagonal representation.

## II. DIRAC EQUATION FOR POWER-LAW POTENTIALS

In the atomic units $\hbar = m = 1$, the three dimensional Dirac Hamiltonian for a four component spinor with "minimal" coupling to the time-independent vector potential $A_\mu = (A_0, \vec{A})$ reads [15]

$$H = \begin{pmatrix} \lambdabar^2 A_0 + 1 & -i\lambdabar\vec{\sigma}\cdot\vec{\nabla} + \lambdabar^2\vec{\sigma}\cdot\vec{A} \\ -i\lambdabar\vec{\sigma}\cdot\vec{\nabla} + \lambdabar^2\vec{\sigma}\cdot\vec{A} & \lambdabar^2 A_0 - 1 \end{pmatrix}, \qquad (2.1)$$

where $\lambdabar$ is the Compton wavelength $\hbar/mc = c^{-1}$ and $\vec{\sigma}$ are the three 2×2 hermitian Pauli matrices. $H$ is measured in units of the rest mass, $mc^2$. Gauge invariance could be used to eliminate the contribution of the off diagonal term $\lambdabar^2\vec{\sigma}\cdot\vec{A}$ in the Hamiltonian (2.1). However, our choice of coupling will be non-minimal, which is accomplished by the replacement $\lambdabar^2\vec{\sigma}\cdot\vec{A} \to \pm i\lambdabar^2\vec{\sigma}\cdot\vec{A}$, respectively. That is the Hamiltonian (2.1) is replaced by the following

$$H = \begin{pmatrix} \lambdabar^2 A_0 + 1 & -i\lambdabar\vec{\sigma}\cdot\vec{\nabla} + i\lambdabar^2\vec{\sigma}\cdot\vec{A} \\ -i\lambdabar\vec{\sigma}\cdot\vec{\nabla} - i\lambdabar^2\vec{\sigma}\cdot\vec{A} & \lambdabar^2 A_0 - 1 \end{pmatrix}. \qquad (2.2)$$

It should be noted that this type of coupling does not support an interpretation of $(A_0, \vec{A})$ as the electromagnetic potential unless, of course, $\vec{A} = 0$ (e.g., the Coulomb potential).

We impose spherical symmetry and write $(A_0, \vec{A})$ as $[V(r), \frac{1}{\lambdabar}\hat{r}W(r)]$, where $\hat{r}$ is the radial unit vector $\vec{r}/r$. $V(r)$ and $W(r)$ are real radial functions referred to as the "even" and "odd" components of the relativistic potential, respectively. Because of spherical symmetry the angular variables could be separated and we can write the radial Dirac equation $(H - \varepsilon)|\chi\rangle = 0$ as [15-17]

$$\begin{pmatrix} \lambdabar^2 V(r) + 1 - \varepsilon & \lambdabar\left[\frac{\kappa}{r} + W(r) - \frac{d}{dr}\right] \\ \lambdabar\left[\frac{\kappa}{r} + W(r) + \frac{d}{dr}\right] & \lambdabar^2 V(r) - 1 - \varepsilon \end{pmatrix} \begin{pmatrix} \varphi^+(r) \\ \varphi^-(r) \end{pmatrix} = 0, \qquad (2.3)$$

where $\kappa$ is the spin-orbit quantum number defined as $\kappa = \pm(j + \tfrac{1}{2}) = \pm 1, \pm 2, \ldots$ for $\ell = j \pm \tfrac{1}{2}$ and $\varepsilon$ is the relativistic energy which is measured in units of $mc^2$. $\varphi^\pm(r)$ are the two components of the radial spinor wave function $\chi(r)$. Examples of relativistic problems that are formulated and solved using this approach are [17]:

1) Dirac-Coulomb: $V(r) = \eta/r$, $W(r) = 0$
2) Dirac-Oscillator: $V(r) = 0$, $W(r) = \eta^2 r$
3) S-wave Dirac-Morse: $\kappa = -1$, $V(r) = Ae^{-\eta r}$, $W(r) = Be^{-\eta r} + \frac{1}{r}$
4) S-wave Dirac-Pöschl-Teller: $\kappa = -1$, $V(r) = 0$, $W(r) = A\tanh(\eta r) + \frac{1}{r}$
5) S-wave Dirac-Hulthén: $\kappa = -1$, $V(r) = A(e^{\eta r} - 1)^{-1}$, $W(r) = B(e^{\eta r} - 1)^{-1} + \frac{1}{r}$

where $A$ and $\eta$ are the physical parameters associated with the corresponding problem and $B^2 = \eta^2 + \lambdabar^2 A^2$. For our present problem the even component of the potential vanishes while the odd component takes the form of the power-law potential $W(r) = A/r^\mu$, where $A$ and $\mu$ are non-zero physical parameters. Therefore, the radial Dirac equation becomes

$$\begin{pmatrix} 1-\varepsilon & \lambdabar\left(\frac{\kappa}{r}+\frac{A}{r^\mu}-\frac{d}{dr}\right) \\ \lambdabar\left(\frac{\kappa}{r}+\frac{A}{r^\mu}+\frac{d}{dr}\right) & -1-\varepsilon \end{pmatrix} \begin{pmatrix} \varphi^+(r) \\ \varphi^-(r) \end{pmatrix} = 0. \qquad (2.4)$$

This equation gives one spinor component in terms of the other as follows:

$$\varphi^\mp(r) = \frac{\lambdabar}{\varepsilon \pm 1}\left(\frac{\kappa}{r}+\frac{A}{r^\mu}\pm\frac{d}{dr}\right)\varphi^\pm(r), \qquad (2.5)$$

where $\varepsilon \neq \mp 1$, respectively. This equation is referred to as the "kinetic balance" relation. Eliminating one spinor component in favor of the other gives the following second order Schrödinger-type differential equation

$$\left[-\frac{d^2}{dr^2}+\frac{\kappa(\kappa\pm 1)}{r^2}+\frac{A^2}{r^{2\mu}}+\frac{A(2\kappa\pm\mu)}{r^{\mu+1}}-\frac{\varepsilon^2-1}{\lambdabar^2}\right]\phi^\pm(r) = 0. \qquad (2.6)$$

If $\mu = 1$, then the problem corresponds to the case of a free Dirac particle which can easily be seen by the redefinition $\kappa \to \kappa + A$. On the other hand, $\mu = -1$ corresponds to the Dirac-Oscillator problem [18]. Moreover, $\mu = 0$ corresponds to a Dirac-Coulomb type problem with $A = Z/\kappa$, where $Z$ is the charge. Consequently, we dismiss these cases by requiring that $\mu \neq 0, \pm 1$. The nonrelativistic limit ($\lambdabar \to 0, \varepsilon \to 1+\lambdabar^2 E$) of Eq. (2.6) shows that the angular momentum quantum number associated with the spinor component $\varphi^+$ is $\ell = \kappa$ ($\ell = -\kappa - 1$) for positive (negative) values of $\kappa$, respectively. While, the corresponding values for $\varphi^-$ are $\ell = \kappa - 1$ and $\ell = -\kappa$, respectively. In the following section we construct an $L^2$ spinor basis with components $\{\phi_n^\pm\}_{n=0}^\infty$ for the solution space of the problem such that the matrix representation of the Dirac wave operator $H - \varepsilon$ is tridiagonal.

### III. TRIDIAGONAL REPRESENTATION AND THE SOLUTION SPACE

The upper component of the spinor basis function which is square integrable and satisfies the boundary conditions could be written as

$$\phi_n^+(r) = a_n x^\alpha e^{-x/2} L_n^\nu(x), \qquad (3.1)$$

where $\alpha$ and $\nu$ are real parameters and satisfy the conditions that $\alpha > 0$ and $\nu > -1$. $L_n^\nu(x)$ is the orthogonal Laguerre polynomial of order $n$ shown in the Appendix. The coordinate $x$ is defined as $x = (\omega r)^\beta$, where $\omega$ and $\beta$ are non-zero real parameters and $\omega$ positive. The integration measure in terms of the $x$ coordinate is $\frac{1}{\omega|\beta|}x^{-1+1/\beta}dx$ since for $\pm\beta > 0$ we get

$$\int_0^\infty dr = \frac{\pm 1}{\omega\beta}\int_0^\infty x^{-1+1/\beta}dx, \qquad (3.2)$$

respectively. Accordingly, the choice for the normalization constant of the basis element in Eq. (3.1) is taken as

$$a_n = \sqrt{\omega|\beta|\Gamma(n+1)/\Gamma(n+\nu+1)}. \qquad (3.3)$$

The requirement of square integrability of $\phi_n^+(r)$ imposes a stronger condition on the parameter $\alpha$ for negative values of $\beta$, which is that $\alpha > -1/2\beta$. Now, the kinetic balance relation (2.5) suggests that the lower component of the spinor basis is obtained from the upper by $\phi_n^- \sim \lambdabar \left( \frac{\gamma}{r} + \frac{\rho}{r^\mu} + \frac{d}{dr} \right) \phi_n^+$. This could be rewritten as $\phi_n^- \sim \lambdabar x^{-1/\beta} \left( \gamma + \frac{1}{2} \rho x^\xi + x \frac{d}{dx} \right) \phi_n^+$, where $\gamma$ and $\rho$ are real dimensionless parameters and $\xi = (1-\mu)/\beta$. It turns out that the solution of the problem is tractable only for integral values of $\xi$. Additionally, the three-term recursion relation for the Laguerre polynomial (A.2) indicates that the tridiagonal representation is obtained only if $\xi = 1$. Therefore, from now on we take $\beta = 1 - \mu$ and, thus, $\beta \neq 0$, 1 or 2. Consequently, we can write

$$\phi_n^- = \frac{2\lambdabar\omega\tau\beta}{x^{1/\beta}} \left( \gamma + \frac{1}{2} \rho x + x \frac{d}{dx} \right) \phi_n^+, \tag{3.4}$$

where $\tau$ is another dimensionless real parameter. These basis parameters will be determined as we proceed. Substituting the expression for $\phi_n^+$ from (3.1) and using the differential and recursion properties of the Laguerre polynomials shown in the Appendix we obtain the following alternative, but equivalent, forms for the basis element of the lower spinor component

$$\begin{aligned}\phi_n^-(r) = \lambdabar\omega\tau\beta\, a_n x^{\alpha-1/\beta} e^{-x/2} \Big[ &2(\gamma+\alpha-\nu)L_n^\nu(x) \\ &+ (1+\rho)(n+\nu)L_n^{\nu-1}(x) + (1-\rho)(n+1)L_{n+1}^{\nu-1}(x) \Big]\end{aligned} \tag{3.5a}$$

$$\begin{aligned}\phi_n^-(r) = \lambdabar\omega\tau\beta\, a_n x^{\alpha-1/\beta} e^{-x/2} \Big\{ &2(\gamma+\alpha)L_n^\nu(x) \\ &- x\left[ (1-\rho)L_n^{\nu+1}(x) + (1+\rho)L_{n-1}^{\nu+1}(x) \right] \Big\}\end{aligned} \tag{3.5b}$$

$$\begin{aligned}\phi_n^-(r) = \lambdabar\omega\tau\beta\, a_n x^{\alpha-1/\beta} e^{-x/2} \Big\{ &2\left[ \left(\gamma+\alpha-\tfrac{\nu+1}{2}\right) + \rho\left(n+\tfrac{\nu+1}{2}\right)\right]L_n^\nu(x) \\ &-(1+\rho)(n+\nu)L_{n-1}^\nu(x) + (1-\rho)(n+1)L_{n+1}^\nu(x) \Big\}\end{aligned} \tag{3.5c}$$

Depending on the range of values of the physical parameters ($A$, $\mu$, and $\kappa$) the solution space will be spanned by one of these three bases elements along with $\phi_n^+(r)$ of Eq. (3.1). We will refer to each of these representations by the equation number of the corresponding basis. To simplify the solution in the first and second representation we take $\gamma = \nu - \alpha$ in (3.5a) and $\gamma = -\alpha$ in (3.5b), respectively. Meeting the square integrability requirement and satisfying the boundary conditions result in constraints on the basis parameters $\alpha$ and $\nu$ shown in the Table.

**Table**: List of constraints on the basis parameters $\alpha$ and $\nu$ obtained by the requirement that the spinor basis with the components (3.1) and (3.5a,b,c) are square integrable and satisfy the boundary conditions.

| $\phi_n^-(r)$ | | $\beta < 0$ <br> ($\mu > 1$) | $1 > \beta > 0$ <br> ($0 > \mu > 1$) | $2 \neq \beta > 1$ <br> ($-1 \neq \mu < 0$) |
|---|---|---|---|---|
| (3.5a): $\gamma = \nu - \alpha$ | $\nu > 0$ | $\alpha > -1/2\beta$ | $\alpha > 1/\beta$ | $\alpha > 1/\beta$ |
| (3.5b): $\gamma = -\alpha$ | $\nu > -1$ | $\alpha > -1/2\beta$ | $\alpha > -1 + 1/\beta$ | $\alpha > 0$ |
| (3.5c): $\rho = \pm 1$ | $\nu > -1$ | $\alpha > -1/2\beta$ | $\alpha > 1/\beta$ | $\alpha > 1/\beta$ |

In the spinor basis $\left\{\psi_n = \begin{pmatrix} \phi_n^+ \\ \phi_n^- \end{pmatrix}\right\}_{n=0}^{\infty}$, the matrix representation of the Dirac wave operator, $H - \varepsilon$, in (2.4) reads as follows

$$\langle \psi_n | H - \varepsilon | \psi_m \rangle = (1 - \varepsilon)\langle \phi_n^+ | \phi_m^+ \rangle - (1 + \varepsilon - 1/\tau)\langle \phi_n^- | \phi_m^- \rangle$$
$$+ \lambda\omega \left\{ \langle \phi_n^+ | x^{-1/\beta} \left[ \kappa - \beta\gamma + x\left(\tfrac{A}{\omega^\beta} - \tfrac{1}{2}\beta\rho\right) \right] | \phi_m^- \rangle + n \leftrightarrow m \right\} \quad (3.6)$$

where the $n \leftrightarrow m$ symbol means that the term inside the curly brackets is repeated with the indices $n$ and $m$ exchanged. The tridiagonal requirement asserts that the term $\langle \phi_n^+ | \phi_m^+ \rangle$ is compatible with the rest if and only if $\beta = 1$ or 2. However, these values have already been dismissed. Consequently, the first term must be eliminated and the solution of the problem is obtained only for $\varepsilon = +1$ (i.e., for the rest mass energy $mc^2$). The negative energy solution for $\varepsilon = -1$ could similarly be obtained as outlined in Sec. IV below. Detailed analysis of the spinor basis with the combined requirements of (1) square integrability, (2) boundary conditions, and (3) tridiagonal representation gives the following three possibilities:

(3.5a) $\beta\kappa > 0$ and $\kappa \neq -1$ ($\ell \neq 0$):
$$\gamma = \kappa/\beta, \ \alpha = (\kappa + 1)/\beta, \ \nu = (2\kappa + 1)/\beta$$

(3.5b) $\beta\kappa < 0$:
$$\gamma = \kappa/\beta, \ \alpha = -\kappa/\beta, \ \nu = -(2\kappa + 1)/\beta$$

(3.5c) $\rho = \text{sign}(\beta A) = \pm 1$:
$$\nu = 2\alpha - 1 - 1/\beta, \ \omega = |2A/\beta|^{1/\beta}$$

In the following subsections we obtain the $L^2$ series solution of the problem associated with each of these three cases.

### A. Solution in the spinor basis (3.1) and (3.5a)

The two components of the spinor basis functions in (3.1) and (3.5a) could now be rewritten as:

$$\phi_n^+(r) = a_n\, x^{\frac{\kappa+1}{\beta}} e^{-x/2} L_n^{\frac{2\kappa+1}{\beta}}(x), \quad (3.7a)$$

$$\phi_n^-(r) = \lambda\omega\tau\beta\, a_n\, x^{\frac{\kappa}{\beta}} e^{-x/2} \left[ (1+\rho)\left(n + \tfrac{2\kappa+1}{\beta}\right) L_n^{\frac{2\kappa+1}{\beta}-1}(x) + (1-\rho)(n+1) L_{n+1}^{\frac{2\kappa+1}{\beta}-1}(x) \right] \quad (3.7b)$$

where $\beta\kappa > 0$ and $\kappa \neq -1$ (i.e., $\ell \neq 0$). Substituting these into (3.6) with $\varepsilon = +1$ and $\gamma = \kappa/\beta$ and using the orthogonality and recurrence relations of the Laguerre polynomials shown in the Appendix we obtain the following elements of the symmetric tridiagonal matrix representation of the Dirac operator

$$(H-1)_{n,n} = \lambda^2\omega^2\beta\tau \left\{ \left(2n+1+\tfrac{2\kappa+1}{\beta}\right)\left[p(\rho^2+1)+2q\rho\right] + 2\left(\tfrac{2\kappa+1}{\beta}-1\right)(p\rho+q) \right\} \quad (3.8a)$$

$$(H-1)_{n,n-1} = -\lambda^2\omega^2\beta\tau \left[p(\rho^2-1)+2q\rho\right] \sqrt{n\left(n+\tfrac{2\kappa+1}{\beta}\right)}, \quad (3.8b)$$

where we have defined the following quantities:
$$p = \beta(1-2\tau), \ q = \tfrac{A}{\omega^\beta} - \tfrac{1}{2}\beta\rho. \quad (3.9)$$

Therefore, the matrix representation of the "wave equation" $(H-1)|\chi\rangle = 0$, where $|\chi\rangle = \sum_m f_m |\psi_m\rangle$, results in the following three-term recursion relation for the expansion coefficients of the wavefunction

$$\left[\left(2n+1+\tfrac{2\kappa+1}{\beta}\right)(\rho^2+1+2\rho q/p)+2\left(\tfrac{2\kappa+1}{\beta}-1\right)(\rho+q/p)\right]f_n$$
$$-(\rho^2-1+2\rho q/p)\sqrt{n\left(n+\tfrac{2\kappa+1}{\beta}\right)}f_{n-1} \qquad (3.10)$$
$$-(\rho^2-1+2\rho q/p)\sqrt{(n+1)\left(n+1+\tfrac{2\kappa+1}{\beta}\right)}f_{n+1} = 0$$

Defining $g_n = \sqrt{\Gamma\left(n+1+\tfrac{2\kappa+1}{\beta}\right)/\Gamma(n+1)}\, f_n$ and

$$\sigma_\pm = (\rho+q/p)^2 - (q/p)^2 \pm 1,\; \zeta = \left(\tfrac{2\kappa+1}{\beta}-1\right)(\rho+q/p), \qquad (3.11)$$

the recursion relation (3.10) takes the following form

$$2\left[\left(n+\tfrac{\kappa+1/2}{\beta}+\tfrac{1}{2}\right)\tfrac{\sigma_+}{\sigma_-}+\tfrac{\zeta}{\sigma_-}\right]g_n - \left(n+\tfrac{2\kappa+1}{\beta}\right)g_{n-1} - (n+1)g_{n+1} = 0. \qquad (3.12)$$

This bares very close resemblance to the three-term recursion relation (A.8) of the Meixner-Pollaczek polynomials $P_n^\lambda(y,\theta)$ [19] shown in the Appendix. Nevertheless, one should take care in pursuing this resemblance and attach some degree of rigor to the investigation of this analogy. Depending on the values of the physical constants in the problem, the parameters that appear in the recursion relation (3.12) may not fall within the permissible range of parameters that define the polynomial $P_n^\lambda(y,\theta)$ as given by Eq. (A.9). To give a clear illustration of this point we start by simplifying the above expressions which is achieved by imposing the "kinetic balance" relation (2.5) on the basis elements. That is, we require that Eq. (3.4) be identical to the relation (2.5) with $\varepsilon = +1$. This gives

$$\tau = 1/4,\; \gamma = \kappa/\beta,\; \rho = 2A/\beta\omega^\beta, \qquad (3.13)$$

resulting in the following parameter assignments:

$$p = \beta/2,\; q = 0,\; \sigma_\pm = \rho^2 \pm 1,\; \zeta = \rho\left(\tfrac{2\kappa+1}{\beta}-1\right). \qquad (3.14)$$

Substituting these in Eq. (3.12) gives one of two recursion relations depending on the range of values of the parameter $\rho$. For $\rho^2 > 1$ (i.e., $\omega < |2A/\beta|^{1/\beta}$) we obtain:

$$2\left[\left(n+\tfrac{\kappa+1/2}{\beta}+\tfrac{1}{2}\right)\cosh\theta + y\sinh\theta\right]g_n - \left(n+\tfrac{2\kappa+1}{\beta}\right)g_{n-1} - (n+1)g_{n+1} = 0, \qquad (3.15a)$$

where $\theta = \sinh^{-1}\left(\tfrac{2\rho}{\rho^2-1}\right)$ and $y = \tfrac{\kappa+1/2}{\beta} - \tfrac{1}{2}$. However, if $1 > \rho^2 > 0$ (i.e., $\omega > |2A/\beta|^{1/\beta}$), then we obtain

$$2\left[\left(n+\tfrac{\kappa+1/2}{\beta}+\tfrac{1}{2}\right)\cosh\theta - y\sinh\theta\right]g_n + \left(n+\tfrac{2\kappa+1}{\beta}\right)g_{n-1} + (n+1)g_{n+1} = 0. \qquad (3.15b)$$

Now we are in a position to make a proper comparison of (3.15a) and (3.15b) with the recursion relation (A.8) of the Meixner-Pollaczek polynomials. Using the well known relations that $\cosh\theta = \cos i\theta$ and $\sinh\theta = -i\sin i\theta$, we define the "Hyperbolic Meixner-Pollaczek polynomials" as

$$\hat{P}_n^\lambda(y,\theta) \equiv P_n^\lambda(-iy, i\theta) = \tfrac{\Gamma(n+2\lambda)}{\Gamma(n+1)\Gamma(2\lambda)} e^{-n\theta}\, {}_2F_1(-n, \lambda+y; 2\lambda; 1-e^{2\theta}), \qquad (3.16a)$$

which satisfies the following modified three-term recursion relation:

$$2\left[(n+\lambda)\cosh\theta + y\sinh\theta\right]\hat{P}_n^\lambda - (n+2\lambda-1)\hat{P}_{n-1}^\lambda - (n+1)\hat{P}_{n+1}^\lambda = 0 \qquad (3.16b)$$

Then, the respective solutions of the recursion relations (3.15a) and (3.15b) read as follows:

$$g_n = \hat{P}_n^{\frac{\kappa+1/2}{\beta}+\frac{1}{2}}\left(\frac{\kappa+1/2}{\beta}-\frac{1}{2},\theta\right), \quad \text{for } \rho^2 > 1, \tag{3.17a}$$

$$g_n = (-)^n \hat{P}_n^{\frac{\kappa+1/2}{\beta}+\frac{1}{2}}\left(-\frac{\kappa+1/2}{\beta}+\frac{1}{2},\theta\right), \quad \text{for } 1 > \rho^2 > 0. \tag{3.17b}$$

Consequently, with the values of the basis parameters in (3.13) and for $\beta\kappa > 0$ and $\kappa \neq -1$ (i.e., $\ell \neq 0$), the $L^2$ series solution of the problem for $\rho^2 > 1$ is given by

$$\chi^a(r) = N^a \sum_{n=0}^{\infty} \sqrt{\Gamma(n+1)/\Gamma(n+1+\tfrac{2\kappa+1}{\beta})}\; \hat{P}_n^{\frac{\kappa+1/2}{\beta}+\frac{1}{2}}\left(\frac{\kappa+1/2}{\beta}-\frac{1}{2},\theta\right)\psi_n^a(r), \tag{3.18a}$$

where $N^a$ is an overall normalization constant that depends only on the physical parameters of the problem $A$, $\mu$, and $\kappa$ and is fixed once and for all. The upper and lower components of the spinor basis element $\psi_n^a(r)$ are given by Eq. (3.7a) and Eq. (3.7b), respectively. On the other hand, for $1 > \rho^2 > 0$, the corresponding solution is

$$\chi^a(r) = N^a \sum_{n=0}^{\infty} (-)^n \sqrt{\Gamma(n+1)/\Gamma(n+1+\tfrac{2\kappa+1}{\beta})}\; \hat{P}_n^{\frac{\kappa+1/2}{\beta}+\frac{1}{2}}\left(-\frac{\kappa+1/2}{\beta}+\frac{1}{2},\theta\right)\psi_n^a(r). \tag{3.18b}$$

In the following subsection we repeat briefly the same development to obtain the solution of the problem for the case where $\beta\kappa < 0$.

### B. Solution in the spinor basis (3.1) and (3.5b)

We could rewrite the two components of the spinor basis functions in (3.1) and (3.5b) as follows:

$$\phi_n^+(r) = a_n\, x^{-\frac{\kappa}{\beta}} e^{-x/2} L_n^{-\frac{2\kappa+1}{\beta}}(x), \tag{3.19a}$$

$$\phi_n^-(r) = -\lambda\omega\tau\beta\, a_n\, x^{-\frac{\kappa}{\beta}+1} e^{-x/2}\left[(1-\rho)L_n^{-\frac{2\kappa+1}{\beta}+1}(x) + (1+\rho)L_{n-1}^{-\frac{2\kappa+1}{\beta}+1}(x)\right], \tag{3.19b}$$

where $\beta\kappa < 0$. Substituting these into (3.6) with $\varepsilon = +1$ and $\gamma = \kappa/\beta$ and using the properties of the Laguerre polynomials shown in the Appendix we obtain

$$(H-1)_{n,n} = \lambda^2\omega^2\beta\tau\left\{\left(2n+1-\tfrac{2\kappa+1}{\beta}\right)\left[p(\rho^2+1)+2q\rho\right]+2\left(\tfrac{2\kappa+1}{\beta}-1\right)(p\rho+q)\right\} \tag{3.20a}$$

$$(H-1)_{n,n-1} = -\lambda^2\omega^2\beta\tau\left[p(\rho^2-1)+2q\rho\right]\sqrt{n\left(n-\tfrac{2\kappa+1}{\beta}\right)}, \tag{3.20b}$$

where the real parameters $p$ and $q$ are as defined in (3.9) above. Following the same development that started with the matrix elements (3.8) leading to the recursion relation (3.12), we obtain

$$2\left[\left(n-\tfrac{\kappa+1/2}{\beta}+\tfrac{1}{2}\right)\tfrac{\sigma_+}{\sigma_-}+\tfrac{\zeta}{\sigma_-}\right]g_n - \left(n-\tfrac{2\kappa+1}{\beta}\right)g_{n-1} - (n+1)g_{n+1} = 0, \tag{3.21}$$

where $g_n = \sqrt{\Gamma\left(n+1-\tfrac{2\kappa+1}{\beta}\right)/\Gamma(n+1)}\, f_n$, $\sigma_\pm$ and $\zeta$ are as defined above in (3.11). Imposing the "kinetic balance" which resulted in the parameters assignments (3.13) and (3.14) and continuing the same line of development as that above, we finally arrive at the following $L^2$ series solution of the problem for the case where $\beta\kappa < 0$:

$$\chi^b(r) = N^b \sum_{n=0}^{\infty} \sqrt{\Gamma(n+1)/\Gamma(n+1-\tfrac{2\kappa+1}{\beta})}\; \hat{P}_n^{-\frac{\kappa+1/2}{\beta}+\frac{1}{2}}\left(\frac{\kappa+1/2}{\beta}-\frac{1}{2},\theta\right)\psi_n^b(r), \tag{3.22a}$$

for $\rho^2 > 1$ and where $N^b$ is an overall normalization constant. The components of the spinor basis element $\psi_n^b(r)$ are given by Eq. (3.19a) and Eq. (3.19b). Now, in the case where $1 > \rho^2 > 0$, the solution becomes

$$\chi^b(r) = N^b \sum_{n=0}^{\infty} (-)^n \sqrt{\Gamma(n+1)/\Gamma(n+1-\tfrac{2\kappa+1}{\beta})} \; \hat{P}_n^{-\tfrac{\kappa+1/2}{\beta}+\tfrac{1}{2}}\left(-\tfrac{\kappa+1/2}{\beta}+\tfrac{1}{2},\theta\right)\psi_n^b(r). \quad (3.22b)$$

In the following subsection we obtain the third solution that corresponds to the choice (3.5c) which is actually needed only if $\beta\kappa > 0$ and $\kappa = -1$ ($\ell = 0$) or if one chooses to take the parameter $\rho = \pm 1$.

### C. Solution in the spinor basis (3.1) and (3.5c)

The two components of the spinor basis functions in this case are:
$$\phi_n^+(r) = a_n x^\alpha e^{-x/2} L_n^\nu(x), \quad (3.23a)$$

$$\phi_n^-(r) = \lambdabar\omega\tau\beta\, a_n x^{\alpha-1/\beta} e^{-x/2} \left\{\left[(2\gamma+1/\beta)+\rho(2n+\nu+1)\right]L_n^\nu(x) \right.$$
$$\left. -(1+\rho)(n+\nu)L_{n-1}^\nu(x) + (1-\rho)(n+1)L_{n+1}^\nu(x)\right\} \quad (3.23b)$$

where $\nu = 2\alpha - 1 - 1/\beta$, $\rho = \text{sign}(\beta A) = \pm 1$, and $\omega = |2A/\beta|^{1/\beta}$. Moreover, we impose the conditions from the Table that $\alpha > 1/\beta$ for $\beta > 0$ and $\alpha > -1/2\beta$ for $\beta < 0$. Inserting these spinor components into (3.6) with $\varepsilon = +1$ and using the orthogonality and recurrence relations of the Laguerre polynomials we obtain the following elements of the symmetric tridiagonal matrix representation of the Dirac operator

$$(H-1)_{n,n} = 4\lambdabar^2\omega^2\beta\tau\left\{p\left[\left(n+\alpha+\rho\gamma+\tfrac{\rho-1}{2\beta}\right)^2\right.\right.$$
$$\left.\left. +\left(n+\alpha-\tfrac{\rho}{2}-\tfrac{1}{2\beta}\right)^2 - \tfrac{\nu^2}{4}\right] + u\left(n+\alpha+\rho\gamma+\tfrac{\rho-1}{2\beta}\right)\right\} \quad (3.24a)$$

$$(H-1)_{n,n-1} = -4\lambdabar^2\omega^2\beta\tau\left[p\left(n+\alpha+\rho\gamma-\tfrac{\rho+1}{2}+\tfrac{\rho-1}{2\beta}\right)+\tfrac{1}{2}u\right]\sqrt{n(n+\nu)}, \quad (3.24b)$$

where the real parameter $p$ is defined in (3.9) above and $u = \rho(\kappa-\beta\gamma)$. Therefore, the "wave equation" $(H-1)|\chi\rangle = 0$, with $|\chi\rangle = \sum_m f_m|\psi_m\rangle$, results in the following three-term recursion relation for the expansion coefficients of the spinor wavefunction

$$\left[(n+\nu+1)(n+d+1)+n(n+d)-\left(\tfrac{\nu+1}{2}\right)^2+z(z+\rho u/p)\right]f_n$$
$$-(n+d)\sqrt{n(n+\nu)}f_{n-1} - (n+d+1)\sqrt{(n+1)(n+\nu+1)}f_{n+1} = 0 \quad (3.25)$$

where we have defined the following two quantities
$$z = \gamma + 1/2\beta, \quad d = \alpha + \rho\gamma - \tfrac{\rho+1}{2} + \tfrac{\rho-1}{2\beta} + \tfrac{u}{2p}. \quad (3.26)$$

Introducing $h_n = \sqrt{\Gamma(n+1)/\Gamma(n+\nu+1)}\, f_n$, we could rewrite this recursion relation in the following form

$$\left[(n+\nu+1)(n+d+1)+n(n+d)-\left(\tfrac{\nu+1}{2}\right)^2+z(z+\rho u/p)\right]h_n$$
$$-n(n+d)h_{n-1} - (n+\nu+1)(n+d+1)h_{n+1} = 0 \quad (3.27)$$

Comparing this with the three-term recursion relation (A.11) for the continuous dual Hahn orthogonal polynomials $S_n^\lambda(y;a,b)$ [19] shown in the Appendix, we conclude that

$$f_n = \sqrt{\frac{\Gamma(n+\nu+1)}{\Gamma(n+1)}} \, S_n^{\frac{\nu+1}{2}}\left(-iy; \frac{\nu+1}{2}, d+\frac{1-\nu}{2}\right), \tag{3.28}$$

where $y^2 = z(z+\rho u/p)$, and we have introduced a modified continuous dual Hahn polynomial $\hat{S}_n^\lambda(y;a,b)$ which we could define as

$$\hat{S}_n^\lambda(y;a,b) \equiv S_n^\lambda(-iy;a,b) = {}_3F_2\left(\begin{array}{c}-n,\lambda+y,\lambda-y\\ \lambda+a,\lambda+b\end{array}\Big|1\right). \tag{3.29}$$

Imposing the kinetic balance condition, which makes $u = 0$, $\gamma = \kappa/\beta$ and $z = \frac{\kappa+1/2}{\beta}$, results in the following $L^2$ series solution for $\rho = \pm 1$ and $\pm\beta\kappa > 0$

$$\chi_\pm^c(r) = N^c \sum_{n=0}^\infty \sqrt{\Gamma(n+\nu+1)/\Gamma(n+1)} \, \hat{S}_n^{\frac{\nu+1}{2}}\left(y; \frac{\nu+1}{2}, \frac{1\mp 1}{2} \pm \frac{2\kappa+1}{2\beta}\right) \psi_n^c(r), \tag{3.30}$$

and where $y^2 = \left(\frac{\kappa+1/2}{\beta}\right)^2$, $N^c$ is a normalization constant, and the components of the spinor basis element $\psi_n^c(r)$ are given by Eq. (3.23a) and Eq. (3.23b).

### IV. DISCUSSION

It is instructive to show that the limited solution we have previously obtained in [6] is a special case of the general solution constructed here. This is done by reducing the tridiagonal representation to a diagonal one. As explained in the Introduction section, this could simply be accomplished by imposing the conditions (1.3) on the recursion relation (1.2). It should also be noted that diagonalization automatically implies that the basis must satisfy the kinetic balance relation. Now, the three-term recursion relation (3.10) which corresponds to (3.5a) and that which corresponds to (3.5b) could be written collectively as follows

$$\left[\left(2n+1\pm\tfrac{2\kappa+1}{\beta}\right)\sigma_+ + 2\zeta\right]f_n - \sigma_-\sqrt{n\left(n\pm\tfrac{2\kappa+1}{\beta}\right)}f_{n-1}$$
$$-\sigma_-\sqrt{(n+1)\left(n+1\pm\tfrac{2\kappa+1}{\beta}\right)}f_{n+1} = 0 \tag{4.1}$$

for $\pm\beta\kappa > 0$ and where $\sigma_\pm$ and $\zeta$ are given by (3.14). The conditions (1.3) give

$$\rho^2 = 1, \text{ and } \tfrac{2\kappa+1}{\beta}(\rho\pm 1) + 1 - \rho = -2n \tag{4.2}$$

The only possible solution of (4.2) is $n = 0$, $\rho = +1$, and $\beta\kappa < 0$. This fully agrees with our earlier result on page 4561 of [6] with the following correspondence between the parameters: $\beta \to (\nu+\tfrac{1}{2})^{-1}$ and $\omega^\beta \to \lambda^2$.

Finally, we address the negative energy solution for which $\varepsilon = -1$. In this case, the kinetic balance relation (2.5) could only be written as $\varphi^+ = \tfrac{\lambda}{2}\left(-\tfrac{\kappa}{r} - \tfrac{A}{r^\mu} + \tfrac{d}{dr}\right)\varphi^-$. This means that in this case the lower spinor component takes the lead. That is, expression (3.1) will be taken for the lower component of the spinor basis whereas the expressions in the set (3.5) refer to the upper component. Moreover, the solution for the case $\varepsilon = -1$ is obtained from that for $\varepsilon = +1$ by the replacement $A \to -A$, $\kappa \to -\kappa$ and $\phi_n^\pm \to \phi_n^\mp$. As an example, the solution (3.30) in this case becomes

$$\chi_\pm^c(r) = N^c \sum_{n=0}^{\infty} \sqrt{\Gamma(n+\nu+1)/\Gamma(n+1)} \; \hat{S}_n^{\frac{\nu+1}{2}}\left(y; \tfrac{\nu+1}{2}, \tfrac{1 \mp 1}{2} \mp \tfrac{2\kappa-1}{2\beta}\right) \psi_n^c(r) \qquad (4.3)$$

for $\rho = \mp 1$ and $\mp \beta\kappa > 0$, and where $y^2 = \left(\tfrac{\kappa-1/2}{\beta}\right)^2$. The components of this spinor basis, $\psi_n^c(r)$, are obtained from (3.23a) and (3.23b) as

$$\phi_n^+(r) = \hbar\omega\tau\beta \, a_n x^{\alpha-1/\beta} e^{-x/2} \left\{\left[(2\gamma+1/\beta)-\rho(2n+\nu+1)\right] L_n^\nu(x) \right.$$
$$\left. +(\rho-1)(n+\nu)L_{n-1}^\nu(x) + (\rho+1)(n+1)L_{n+1}^\nu(x)\right\} \qquad (4.4a)$$

$$\phi_n^-(r) = a_n x^\alpha e^{-x/2} L_n^\nu(x) \qquad (4.4b)$$

## APPENDIX

The following are useful formulas and relations satisfied by the orthogonal polynomials that are relevant to the developments carried out in this work. They are found on most textbooks and monographs on orthogonal polynomials [19,20]. We list them here for ease of reference. In these formulas $_2F_1$ stands for the hypergeometric function, $_1F_1$ is the confluent hypergeometric function, $_3F_2$ is the generalized hypergeometric series, and $\Gamma$ is the gamma function.

1) The Laguerre polynomials $L_n^\nu(x)$:

$$L_n^\nu(x) = \frac{\Gamma(n+\nu+1)}{\Gamma(n+1)\Gamma(\nu+1)} \, _1F_1(-n;\nu+1;x) \qquad (A.1)$$

where $\nu > -1$ and $n = 0, 1, 2, \ldots$

$$xL_n^\nu = (2n+\nu+1)L_n^\nu - (n+\nu)L_{n-1}^\nu - (n+1)L_{n+1}^\nu \qquad (A.2)$$

$$xL_n^\nu = (n+\nu)L_n^{\nu-1} - (n+1)L_{n+1}^{\nu-1} \qquad (A.3)$$

$$L_n^\nu = L_n^{\nu+1} - L_{n-1}^{\nu+1} \qquad (A.4)$$

$$x\frac{d}{dx}L_n^\nu = nL_n^\nu - (n+\nu)L_{n-1}^\nu \qquad (A.5)$$

$$\left[x\frac{d^2}{dx^2} + (\nu+1-x)\frac{d}{dx} + n\right]L_n^\nu(x) = 0 \qquad (A.6)$$

$$\int_0^\infty x^\nu e^{-x} L_n^\nu(x) L_m^\nu(x) dx = \frac{\Gamma(n+\nu+1)}{\Gamma(n+1)} \delta_{nm} \qquad (A.7)$$

2) The Meixner-Pollaczek polynomials $P_n^\lambda(y,\theta)$:

$$2\left[(n+\lambda)\cos\theta + y\sin\theta\right]P_n^\lambda - (n+2\lambda-1)P_{n-1}^\lambda - (n+1)P_{n+1}^\lambda = 0 \qquad (A.8)$$

$$P_n^\lambda(y,\theta) = \frac{\Gamma(n+2\lambda)}{\Gamma(n+1)\Gamma(2\lambda)} e^{in\theta} \, _2F_1(-n, \lambda+iy; 2\lambda; 1-e^{-2i\theta}) \qquad (A.9)$$

where, $\lambda > 0$ and $0 < \theta < \pi$.

$$\int_{-\infty}^{+\infty} \rho^\lambda(y,\theta) P_n^\lambda(y,\theta) P_m^\lambda(y,\theta) dy = \frac{\Gamma(n+2\lambda)}{\Gamma(n+1)} \delta_{nm} \qquad (A.10)$$

where $\rho^\lambda(y,\theta) = \frac{1}{2\pi}(2\sin\theta)^{2\lambda} e^{(2\theta-\pi)y} |\Gamma(\lambda+iy)|^2$

3) The continuous dual Hahn polynomials $S_n^\lambda(y;a,b)$:

$$y^2 S_n^\lambda = \left[(n+\lambda+a)(n+\lambda+b) + n(n+a+b-1) - \lambda^2\right] S_n^\lambda \qquad (A.11)$$
$$-n(n+a+b-1)S_{n-1}^\lambda - (n+\lambda+a)(n+\lambda+b)S_{n+1}^\lambda$$

$$S_n^\lambda(y;a,b) = {}_3F_2\left(\begin{array}{c}-n,\lambda+iy,\lambda-iy\\ \lambda+a,\lambda+b\end{array}\bigg|1\right) \qquad (A.12)$$

where $y^2 > 0$ and $\lambda$, $a$, $b$ are positive except for a pair of complex conjugates with positive real parts.

$$\int_0^\infty \rho^\lambda(y) S_n^\lambda(y;a,b) S_m^\lambda(y;a,b) dy = \frac{\Gamma(n+1)\Gamma(n+a+b)}{\Gamma(n+\lambda+a)\Gamma(n+\lambda+b)}\delta_{nm} \qquad (A.13)$$

where $\rho^\lambda(y) = \frac{1}{2\pi}\left|\frac{\Gamma(\lambda+iy)\Gamma(a+iy)\Gamma(b+iy)}{\Gamma(\lambda+a)\Gamma(\lambda+b)\Gamma(2iy)}\right|^2$.